\documentclass[aps,tightenlines,twocolumn,superscriptaddress]{revtex4-1}
\usepackage{graphicx}
\usepackage{dcolumn}
\usepackage{bm}
\usepackage{csquotes}
\usepackage{epigraph}
\usepackage{amssymb}
\usepackage{color,graphicx}
\usepackage{subfigure}
\usepackage{lmodern}
\usepackage{amsmath}
\usepackage{amsbsy}
\usepackage{amsthm}
\usepackage{float}
\usepackage{bbm}

\begin{document}

\title{Quantum coherence on selectivity and transport of ion channels}

\author{Mina Seifi}
\affiliation{Research Group on Foundations of Quantum Theory and Information,
Department of Chemistry, Sharif University of Technology
P.O.Box 11365-9516, Tehran, Iran}
\author{Ali Soltanmanesh}
\affiliation{Research Group on Foundations of Quantum Theory and Information,
Department of Chemistry, Sharif University of Technology
P.O.Box 11365-9516, Tehran, Iran}
\affiliation{Sharif Quantum Center, Sharif University of Technology, Tehran, Iran}
\author{Afshin Shafiee*}
\affiliation{Research Group on Foundations of Quantum Theory and Information,
Department of Chemistry, Sharif University of Technology
P.O.Box 11365-9516, Tehran, Iran}
\begin{abstract}
Recently, it has been suggested that ion channel selectivity filter may exhibit quantum coherence, which may be appropriate to explain ion selection and conduction processes. Potassium channels play a vital role in many physiological processes. One of their main physiological functions is the efficient and highly selective transfer of \text{K}$^{+}$ ions through the membranes into the cells. To do this, ion channels must be highly selective, allowing only certain ions to pass through the membrane,  while preventing the others. The present research is an attempt to investigate the relationship between hopping rate and maintaining coherence in ion channels. Using the Lindblad equation to describe a three-level system, the results in different quantum regimes are examined. We studied the distillable coherence and the second order coherence function of the system. The oscillation of distillable coherence from zero, after the decoherence time, and also the behavior of the coherence function clearly show the point that the system is coherent in ion channels with high throughput rates.
\end{abstract}

\maketitle

\section{Introduction}

\begin{figure}
\centering
\includegraphics[scale=0.36]{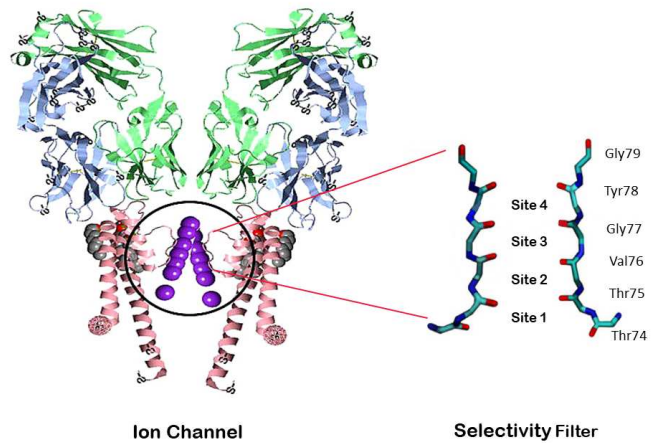}
\caption{(Left) A representation of KcsA ion channel after PDB 1K4C. (Right) Two P-loop monomers in the selectivity filter.}
\label{ion}
\end{figure}

 Quantum biology is a relatively new field of study in quantum mechanics which can use quantum theory in some aspects of biology that classical physics cannot describe  precisely. In the beginning, it was believed that quantum phenomena such as tunneling or quantum entanglement do not exist in living environments since these environments are inherently warm, humid, and noisy \cite{Moh,Tus,Bne}. The newfound evidence, in recent years, reveals that quantum principles play a critical role in explaining various biological phenomena such as photosynthesis, quantum effects in the brain, and spin and electromagnetic routing of the migratory birds \cite{ghas,Mar,arash,Sch,kim,Lam}. According to this evidence, the role of quantum phenomena, such as tunneling and quantum coherence, has been widely accepted in the crucial activities of living cells \cite{ghasem}. Recently, it has been proposed that quantum coherence may play a role in the selectivity of ions and their transport through ion channels \cite{Gan,sal,vaz}. Due to the energy scale and transport phenomena, ion channels can be considered as a distinct protein system that the quantum effects may have a functional role within them so that their activities can be comprehended via quantum mechanics \cite{Gan}. These channels are a collection of proteins embedded in the cell membrane that tune the flux of specific ions across the membrane and regulate interactions between the cell and its environment \cite{Cor}. Structurally, ion channels are protein complexes comprised of several subunits whose cyclic arrangement forms sub-nanometer pores for ions to enter or leave the cell \cite{Moh}. Ion channels share common properties, the most important of which is the presence of a gate  that can be activated by such factors as chemicals, voltage, light, and mechanical pressure. Another common feature is having a selectivity filter responsible for passing only one specific type of ion. The structure of this filter is well studied in the Streptomyces lividans (KcsA) bacterial channel \cite{doyl}. The 3.4 nm long KcsA channel is comprised of a 1.2 nm long selectivity filter that is composed of four P-loop monomers. Each P-loop is composed of five amino acids: (Theronine (Thr75), Valine (Val76), Glycine (Gly77), Tyrosine (Tyr78), Glycine (Gly79)) linked by peptide units (H-N-C=O). 
The selectivity filter width is only a few angstroms (3A). The ions must move in a single file without their hydration shell in this filter. The  process of gating, i.e., the mechanism that controls the closing and opening of the pores, is different in the variety of potassium channels, but the sequence of amino acids forming the selection filter is the same in all potassium channels \cite{doyl}. The selectivity filter is capable of selecting potassium ions over sodium ions in a ratio of $10^{4}$ : 1.   Ion channels allow ions to enter or leave cells in a very selective and rapid manner. FIG. \ref{ion} shows the structure of the KcsA channel and its selectivity filter. Potassium channels conduct $\text{K}^{+}$ ions at a rate of $10^{6}$-$10^{8}$ $\text{s}^{-1}$ throughout the cell membrane \cite{tri,mor}. An important question is how a flexible structure, such as a selectivity filter, can be selected at high speed. This rapid and high selectivity is critical for the physiology of living creatures. Various ion channels are involved in several biological processes, including nerve signaling, muscular contraction, cellular homeostasis, and epithelial fluid transport \cite{wes,ogr,par}. The selectivity of ion channels refers to the fact that each ion channel is individually for passing the specific ions. For example, potassium channels only allow the potassium ions to pass through the membrane while rejecting the other ions (e.g., sodium ions) \cite{wes}. The ion radii of sodium and potassium ions only differ by 0.38A. Despite this slight difference in ion radii, the selectivity of the potassium ions is more than a thousand times higher than that of sodium ions \cite{mac}. Hence, the high selectivity of the ion channels cannot be simply explained by the physical obstruction, and the other factors may be involved \cite{Ash}. There is a large body of literature and numerous hypotheses about ion selectivity in biomolecular and genetic disciplines. Many researchers have contributed to developing the current views and concepts by experimentation and simulation \cite{tho,kas,dud}. Qi \textit{et al.} examined the importance of channel size in ion transport selectivity in molecular detail \cite{Qi}. To investigate the selectivity and other properties of the ion channel, Allen \textit{et al.} performed molecular dynamic calculations on the entire experimental protein structure determined for the KcsA potassium channel from Streptomyces lividans \cite{allen}. Allen and Chung used three-dimensional Brownian dynamics simulations to study the conductivity of the KcsA potassium channel using a known crystallographic structure \cite{chung}. Salari \textit{et al.} investigated the possibility of quantum ion interference through ion channels to understand the role of quantum interference on selectivity in ion channels \cite{sal}. Summhammer \textit{et al.} also analyzed the solutions of the Schroedinger equation for the bacterial KcsA ion channel. They claimed that quantum mechanical calculation is needed to explain basic biological properties such as ion selection in the ion channels of the trans membrane \cite{summ}. Despite the extensive experimental and theoretical researches, most biomolecular methods cannot well explain the ion selection in small scale (nanoscale), and still, numerous issues remain unresolved.  After determining the three-dimensional structure of the bacterial KcsA channel with atomic resolution using x-ray crystallography, doubts were raised about the classical explanations, such as sung fit. It seems that such a process cannot well explain ion selectivity \cite {nos,jia,rou}. Considering the inadequacy of classical mechanics in explaining ion selectivity as well as experimental evidences demonstrating the existence of quantum effects in biological cells, it seems that quantum mechanics should be used to resolve this problem. To comprehend the ion selection as well as ion transport, accurate atomic models which are able to well describe the microscopic interactions are required. The coherence functional relations can only be guessed in the ion channels. In other words, coherence may play a role in ion selectivity. Vaziri \textit{et al.} suggested that the ion channel selectivity filter shows quantum coherence, which could possibly explain the ion selection process \cite{vaz}.\\In the present work, the relationship between hopping rate and maintaining coherence in ion channels is investigated. The Lindblad master equation  is used to describe the system, and the coherence of the system  is examined at different hopping rates. Selectivity and high rates in ion channels seem to be two contradictory features. We show that these two features are not going to be contradictory to each other, but high rate is an essential requisite for selectivity. High selectivity occurs only at high hopping rates. Despite the occurrence of decoherence, the high rate of hopping causes the system to always maintain its coherence in an oscillating manner.\\This paper is organized as follows. In Sec. II, we briefly review the Lindblad master equation and decoherence in open quantum systems. In Sec. III, the quantum mechanical model for  ion transition through the ion channel is presented, and quantum transport equations in terms of Lindblad operators are introduced. Then, the results are discussed and the effect of hopping rates on coherence is investigated in detail in Sec. IV. Finally,  the paper is concluded in Sec. V.  

\section{Decoherence and Lindblad master equation}

All realistic systems inevitably interact with their environments. For the quantum systems in nanoscale and quantum biological systems, these interactions are not negligible; therefore, they should be considered as open systems \cite {Zha}. In recent years, understanding the dynamics of open systems has been one of the most complex challenges in quantum physics. The environment has an important role in the quantum realm. The interaction of a system and its environment can result in the entanglement between the system and environment in the quantum realm so that the system entity may generally change \cite {sch}. In this interaction, the environment has an infinite degree of freedom, and the system cannot be characterized with wave function or specific states. So, a density matrix should be assigned to the central system, and the evolution of this matrix is important over time \cite {sch}. The terms associated with the quantum behavior emerge as diagonal elements of the density matrix operator. Due to the system and environment interactions, quantum coherence which is characterized by the non-diagonal elements decay, and quasi-classical features occur in the system. Based on decoherence, if the interactions of a quantum system and its environment are considered, a realistic image of a distinct model can be achieved \cite{tir}. The complexity of the transmission dynamics in ion channels is attributed to the high interactions between ions along with the high degree of freedom of the environment. The selectivity filter is responsible for the passage of certain ions in ion channels. There are three sites for this filter, as shown in FIG. \ref{ion}. Jumping from one site to another is accompanied by an energy barrier. The jump rate between these sites is called the hopping rate. The rate of hopping between sites should be equal to the rate of transmission($10^{6}$-$10^{8}$\text{s}$^{-1}$) through the channel. To maintain the quantum state of the ions during the passage of the ion channel, the decoherence time must be greater than the time interval of the ion passing through the channel (10-20 nanoseconds) \cite{vaz}. Unlike a closed system, the temporal evolution of an open system is non-unitary due to the dissipative terms in the master equation. The Lindblad master equation is of special importance because it is the most general generator of Markovian dynamics in quantum systems \cite{man,dub}.  
Here, the Lindblad master equation can be written in a generic state as follows \cite{qin,solta,nae}:
\begin{equation}
\label{lindblad} 
\frac{d}{dt}\hat{\rho}(t)=-i[\hat{H}(t),\hat{\rho}(t)]+\sum_i\gamma(\nu_i\rho_t\nu^{\dagger}_i-\frac{1}{2}\nu^{\dagger}_i\nu_i\rho_t-\frac{1}{2}\rho_t\nu^{\dagger}_i\nu_i)
\end{equation}  
The first term on the right side of the above equation is the Liouvillian part which represents the unitary evolution of the density matrix.  The second term is Lindbladian which illustrates the system's interaction with its environment and represents decoherence effects \cite{ale}. The Lindblad master equation is local and markovian over time. The $ \nu$ operators are not necessarily hermitic and are known as Kraus operators, satisfying ${\sum_L\hat{k}_{L}^\dagger \hat{k}_{L}}=1$. A simple application of this equation in quantum optics is photon emission from a two-level atom in free space. In this situation, the density matrix is transformed into a 2×2 matrix, and the $\nu$ operators are reduced to Pauli lowering and rising operators. Consequently, the Lindblad equation is simply transformed into four first-order linear differential equations \cite{che}.

Equation (1) can be symbolically expressed as follows:
\begin{equation}
\label{symblindblad}
\frac{d}{dt}\hat{\rho}(t)=L^{\dagger}_i\hat{\rho}(t)
\end{equation}
where L is a Lindbladian super-operator that is implemented on the density matrix and leads to its temporal evolution. Since $L^{\dagger}_i$ is a linear map in the operator space, it is referred to as a super-operator. In the present work, the evolution of the reduced density matrix is investigated using the Lindblad master equation. In the following, the employed model is described in detail.
 

\section{Model and Methods}

\begin{figure}
\centering
\includegraphics[scale=0.35]{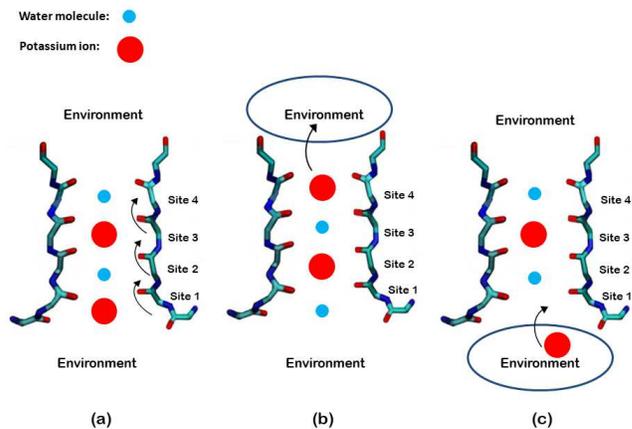}
\caption{Schematic illustration of the ion channel states corresponds to (a)$=\vert 0\rangle$ (b)$=\vert 1\rangle$ (c)$=\vert 2\rangle$ states. }
\label{filter}
\end{figure}

Herein, we express our model to discuss the high speed of selectivity in ion channels using Lindblad master equation. The mechanism of ion permeation through the selectivity filter in ion channels has been investigated using diverse experimental techniques such as radiotracer flux assays, single-channel electrophysiological measurements, x-ray crystallography, and molecular dynamics (MD) simulations. These experiments support two mechanisms commonly referred to as “knock–on” (water molecules may or may not be present in this mechanism) and “hard-knock” (water molecules are ignored in it) permeation models. To determine whether these experimental features can be explained by either of the two permeation models, Huong T. Kratochvil et al. conducted molecular dynamics (MD) simulations and computed 2D IR spectra for all relevant ion configurations \cite{kra}. They showed that knock-on model with water molecules is in most agreement with experience. Accordingly, we considered the presence of water molecules in our model. We examined our model on a relatively short time scale. Due to the large fluctuations of ions in the ion channel, there is probably no strong columbine repulsion at this time scale. To this end, an ion channel with four sites is considered \cite{kra} (FIG. \ref{filter}). In this model we consider a three state system: 1) Potassium ions in the first and third sites and water molecules in the second and fourth sites. 2) Potassium ions in the second and fourth states with water molecules in the first and third sites. 3) A Potassium ion releases to the environment from the previous state. The jump from site 4 to the environment is considered as a 3 to 1 transition. Jumping between these sites is associated with an energy barrier. Due to the high speed of passing, there is always a state transition from site 1 to 2, 2 to 3, or 3 to 1. 
The system can be shown in a three-dimensional Hilbert space with three states of $\vert 0\rangle$ , $\vert 1\rangle$ and $\vert 2\rangle$ similar to a spin-1 system.

The evolution of the density matrix is obtained using the Lindblad master equation as below ($\hslash$ is assumed to be one in the rest of the paper):
\begin{equation}
\label{lindblad} 
\frac{d}{dt}\hat{\rho}(t)=-i[\hat{H}(t),\hat{\rho}(t)]+L[\hat{\rho}(t)]
\end{equation} 
Also, the time-dependent Hamiltonian is considered as: \begin{equation}
\label{totalH}
\hat{H}=\hat{H}_0+\hat{H}_1
\end{equation}
\begin{align}
\label{H0}
\hat{H}_0=\omega_0\hat{S}_Z
\end{align}
\begin{equation}
\label{H1} 
\hat{H}_1=c(\vert 1\rangle\langle 0\vert +\vert 2\rangle\langle 1\vert +\vert 0\rangle\langle 2\vert)
\end{equation} 
where $\hat{H}_0$  is the Hamiltonian system with a transition frequency $\omega_0$, $ \hat{S}_Z $ is the z-coordination operator of spin-1 and $\hat{H}_1$ is  associated with the coupling of the system and its environment. Here, c is the hopping rate coefficient, and L is a super-operator: 
\begin{equation}
\label{superoperator}
L[\hat{\rho}]=\gamma(\hat{S}_-\hat{\rho}\hat{S}_+ -\frac{1}{2}\hat{S}_+\hat{S}_-\hat{\rho} -\frac{1}{2}\hat{\rho}\hat{S}_+\hat{S}_-)
\end{equation}
where $\hat{S}_\pm=\dfrac{1}{\sqrt{2}}(\hat{S}_x \pm i\hat{S}_y )$ are the spin-1 operators. \\The Hamilton function, $\hat{H}_1$,  expresses the transition between states by passing particles from one site to the next. To drive the corresponding equations in the Dirac picture, the system density matrix and the Hamiltonian interaction are rewritten as follow: 
\begin{equation}
\label{density}
\hat{\rho}^{D}(t)=e^{it\hat{H}_0}\hat{\rho}(t)e^{-it\hat{H}_0}
\end{equation}
\begin{equation}
\label{interaction}
\hat{H}^{D}_{1} (t)=e^{it\hat{H}_0}\hat{H}_1e^{-it\hat{H}_0}
\end{equation}
Using Equations \eqref{density} and \eqref{interaction} and going to the Dirac picture, the Lindblad equation becomes: 
\begin{equation}
\label{dirac} 
\frac{d}{dt}\hat{\rho}^{D}(t)=-i[\hat{H}^{D}_{1}(t),\hat{\rho}^{D}(t)]+L^{D}[\hat{\rho}^{D}(t)]
\end{equation} 
where 
\begin{equation}
\label{superoperatordirac}
L^{D}[\hat{\rho}^{D}]=\gamma(\hat{S}_-\hat{\rho}^{D}\hat{S}_+ -\frac{1}{2}\hat{S}_+\hat{S}_-\hat{\rho}^{D} -\frac{1}{2}\hat{\rho}^{D}\hat{S}_+\hat{S}_-)
\end{equation}
According to the above discussion, equation \eqref{dirac} can be rewritten  in the basis of the eigenstates of $ \hat{S}_z $  operator  as follows:
\begin{align}
\label{num}
 \nonumber
\frac{d}{dt} \hat{\rho}^{D_{00}}&=-ic(\hat{\rho}^{D_{20}}-\hat{\rho}^{D_{01}})-2\gamma\hat{\rho}^{D_{00}}\\ \nonumber
\frac{d}{dt}\hat{\rho}^{D_{01}}&=e^{i\omega_{0}t}(ic\hat{\rho}^{D_{21}}+ic\hat{\rho}^{D_{02}}-2\gamma\hat{\rho}^{D_{01}})\\
 \nonumber \frac{d}{dt}\hat{\rho}^{D_{02}}&=e^{2i\omega_{0}t}(-ic\hat{\rho}^{D_{22}}+ic\hat{\rho}^{D_{00}}-\gamma\hat{\rho}^{D_{02}})\\
 \nonumber \frac{d}{dt}\hat{\rho}^{D_{10}}&=e^{-i\omega_{0}t}(-ic\hat{\rho}^{D_{00}}+ic\hat{\rho}^{D_{11}}-2\gamma\hat{\rho}^{D_{10}})\\
\frac{d}{dt}\hat{\rho}^{D_{11}}&=-ic(\hat{\rho}^{D_{01}}-\hat{\rho}^{D_{12}})+2\gamma(\hat{\rho}^{D_{00}}-\hat{\rho}^{D_{11}})\\
 \nonumber \frac{d}{dt}\hat{\rho}^{D_{12}}&=-ice^{i\omega_{0}t}(\hat{\rho}^{D_{02}}-\hat{\rho}^{D_{10}})+\gamma e^{i\omega_{0}t}(2\hat{\rho}^{D_{01}}-\hat{\rho}^{D_{12}})\\
 \nonumber \frac{d}{dt}\hat{\rho}^{D_{20}}&=-ice^{-2i\omega_{0}t}(\hat{\rho}^{D_{10}}-\hat{\rho}^{D_{21}})-\gamma e^{-2i\omega_{0}t}\hat{\rho}^{D_{20}}\\
 \nonumber \frac{d}{dt}\hat{\rho}^{D_{21}}&=-ice^{-i\omega_{0}t}(\hat{\rho}^{D_{11}}-\hat{\rho}^{D_{22}})+\gamma e^{-i\omega_{0}t}(2\hat{\rho}^{D_{10}}-\hat{\rho}^{D_{21}})\\
 \nonumber\frac{d}{dt}\hat{\rho}^{D_{22}}&=-ic(\hat{\rho}^{D_{12}}-\hat{\rho}^{D_{21}})+2\gamma\hat{\rho}^{D_{11}}
\end{align}
\section{Results and Discussion}
\begin{center}
\begin{figure*}
\centering
\subfigure[]{\label{prob1}
\includegraphics[scale=0.4]{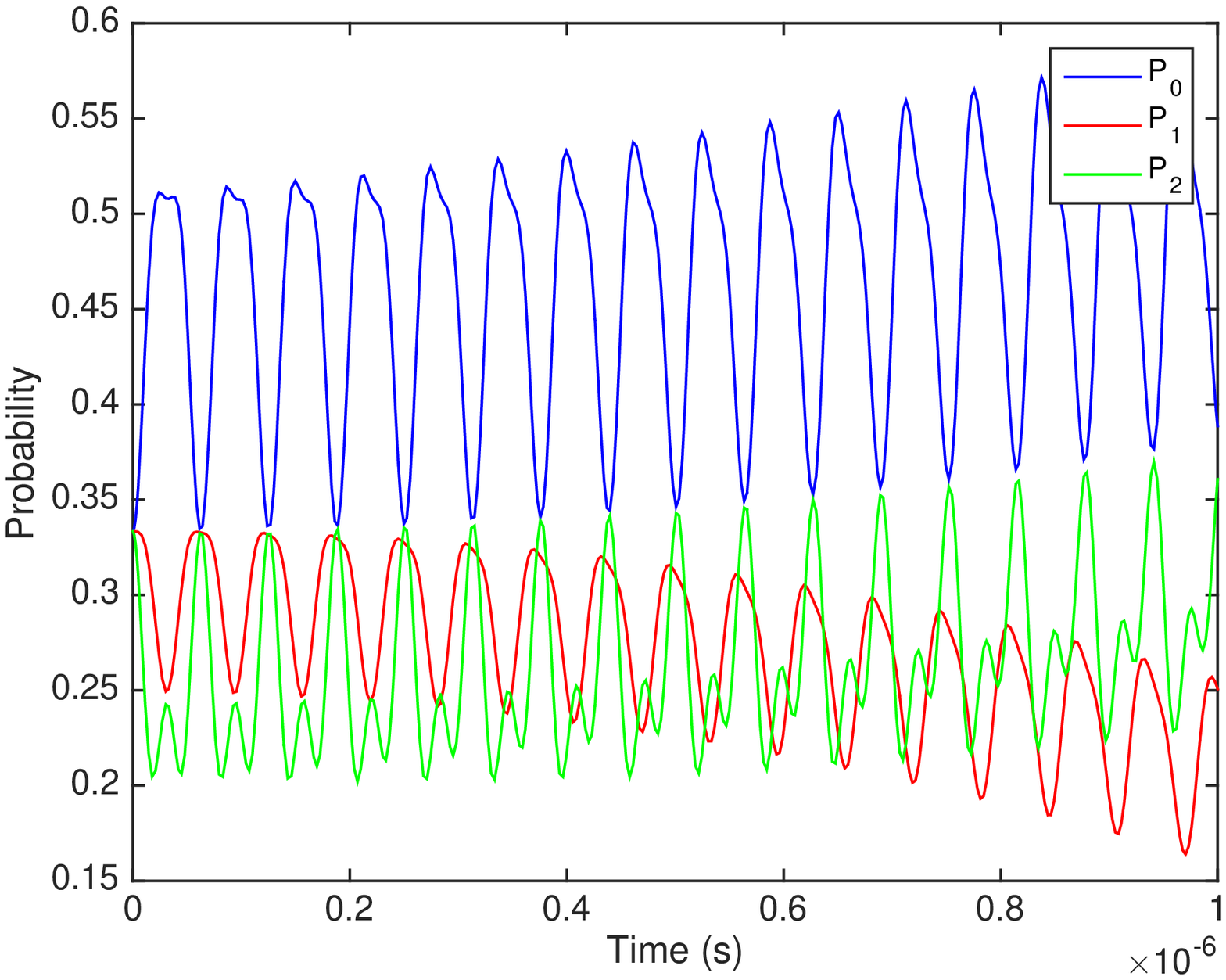}}
\subfigure[]{\label{prob2}
\includegraphics[scale=0.4]{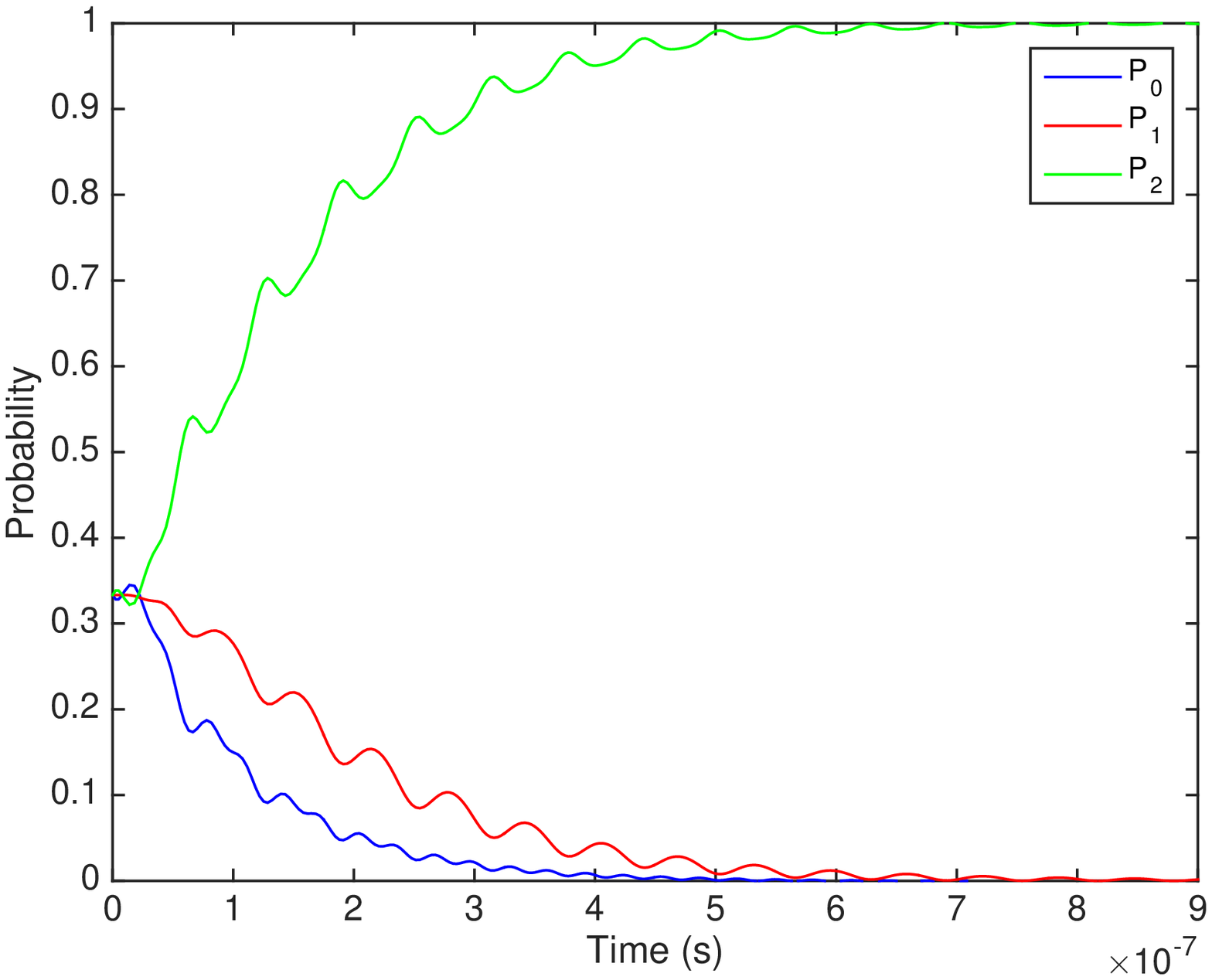}}
\caption{The probability of finding the system in $|0\rangle, |1\rangle$ and $|2\rangle$ states versus time, with $\omega=1\times 10^8s^{-1}$ and (a) $\gamma=0$ and $c=2\times 10^7s^{-1}$ (b) $\gamma=0.5\times10^7s^{-1}$ and $c=2\gamma$.}
\end{figure*}
\end{center}

The mechanism behind the high throughput rate in $\text{K}^+$ channels and its contradiction with its high selectivity is still an open problem. However, it has been suggested that quantum coherent hopping is included throughout the process \cite{Gan,vaz,sala,bhat}. Nevertheless, the system is coupled to a biological environment, which causes the system to go through a decoherence process. The mentioned problem is a tremendous obstacle in understanding the mechanism in the quantum regime. To study the problem, first we solved the system of equations \eqref{num} with the system initial state being in a superposition:
\begin{align}
\label{is}
\vert\psi_i\rangle=\frac{1}{\sqrt{3}}(|0\rangle+|1\rangle+|2\rangle).
\end{align}

\begin{center}
\begin{figure*}
\centering
\subfigure[]{\label{real}
\includegraphics[scale=0.4]{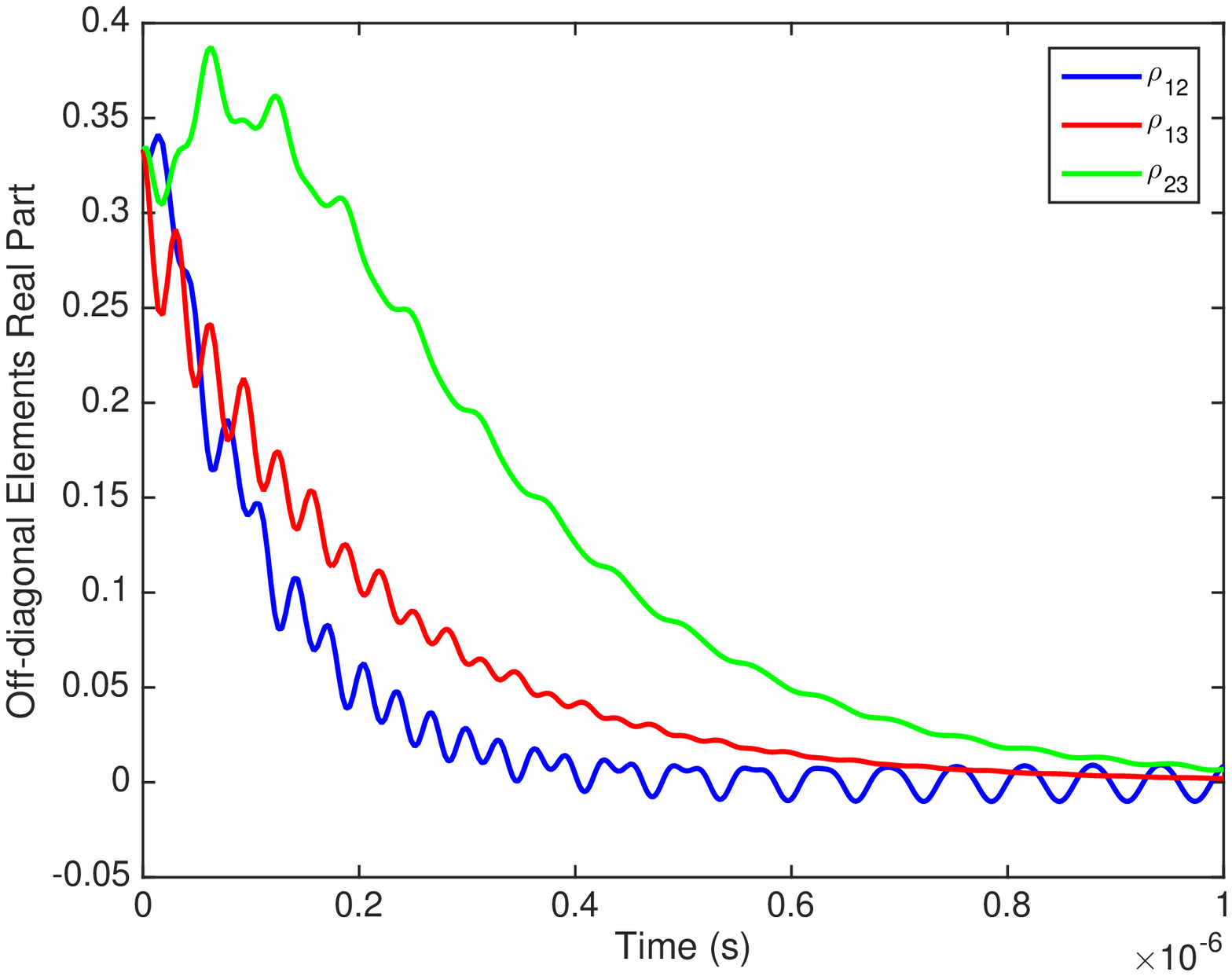}}
\subfigure[]{\label{imag}
\includegraphics[scale=0.4]{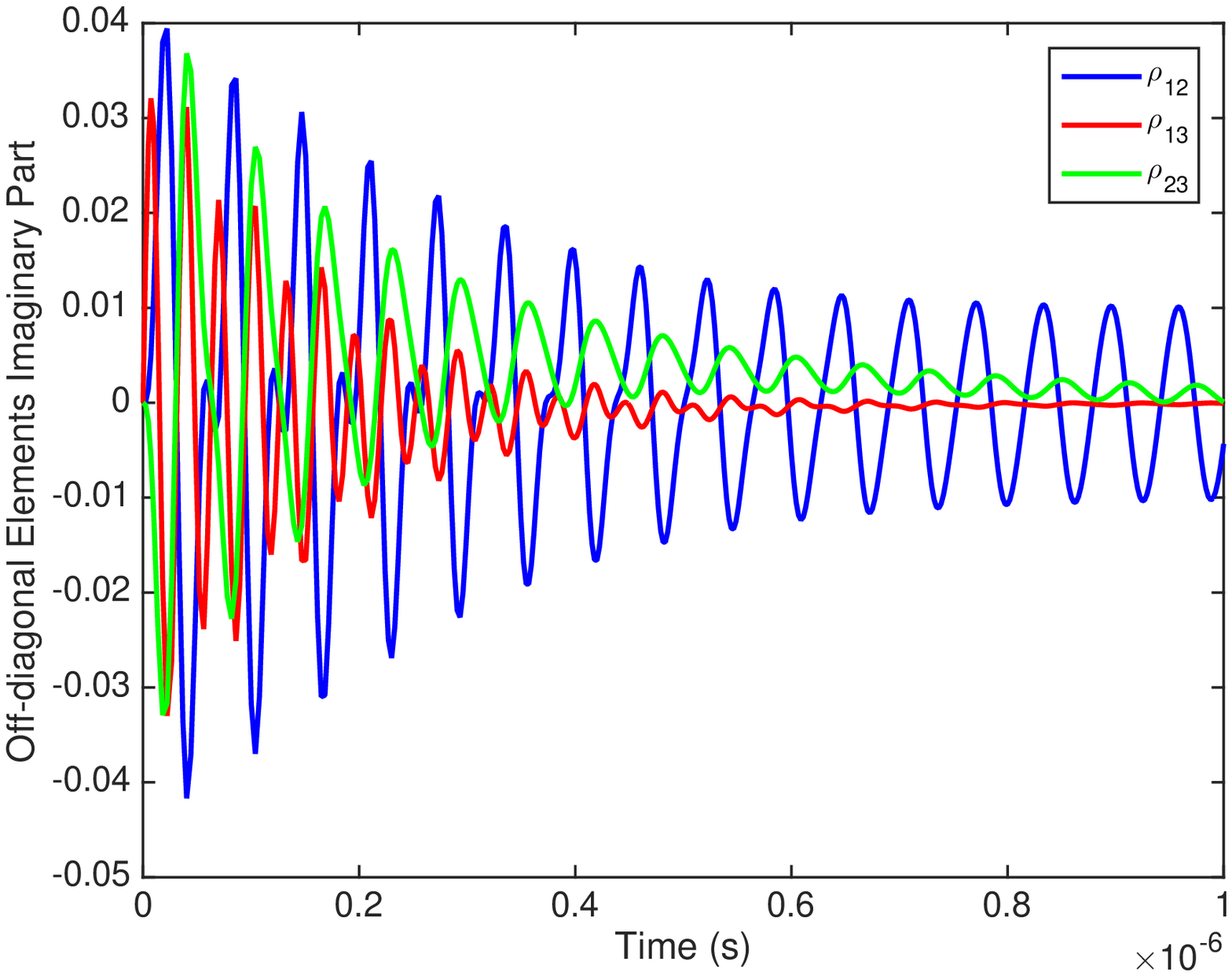}}
\caption{The evolution of (a) real parts and (b) imaginary parts, of off-diagonal elements of density matrix versus time. With $\gamma=0.5\times10^7s^{-1}$ and $c=2\gamma$.}
\end{figure*}
\end{center}

In the absence of the environment (in a case with $\gamma=0$) the probabilities of the states oscillate with time as we expect. However, the probability of finding the system in the $|1\rangle$ state tends to zero as time passes by (see Fig.~\ref{prob1}). This phenomenon shows us that $|1\rangle$ is just an intermediary state with a short lifetime. The use of the spin 1 matrix in the Hamiltonian system indicates that these states are not degenerative. The difference in energy of these states causes them to behave differently. In the spin model, the middle term is considered as an intermediary state. We also considered its energy in proportion to the intermediary state. It is expected that the system will be found more likely in the states where a potassium ion enters the channel from the environment or enters the environment from the channel($|0\rangle$ and $|2\rangle$ respectively). On the other hand, it is found less likely in the intermediary state. This expectation is consistent with our results.
 Otherwise, in a case in which decoherence is present (with $\gamma>0$) the probabilities oscillate and tend to a finite value after the decoherence time, as shown in Fig.~\ref{prob2}. As the values of$\gamma$ increases, makes the system to equilibrate faster. The appropriate gamma for our system varies between  $\gamma=0.5\times10^{7}s^{-1} $ and $2\times10^{8}s^{-1}$ \cite{vaz}. The larger the number in this range, the more fluctuations we see. After the decoherence time we can almost find the system with the lowest energy, in state $|2\rangle$.  In other words, the system prefers to cross and release the particles to the environment quickly.
Also, Fig. \ref{real} and \ref{imag} shows a study on the evolution of off-diagonal elements of the density matrix of the system. Surprisingly, the off-diagonal elements tends to zero as time passes by. But, it seems the fast hopping speed of particles prevents the off-diagonal elements to became zero, and some oscillation can be observed even after the decoherence time. This behavior encourages us to study the coherence of the system. If the selectivity process considered to be quantum mechanical, it is required for the system to remain coherent.

\begin{table}[!t]
\centering
\caption{Requirements for a coherence measure $C_d(\hat{\varrho})$. $S$ is von Neumann entropy and $\Delta$ represents the dephasing operator.  \label{table1}}
\begin{tabular}{p{18ex}  p{32ex}}
\hline
Postulate & Definition \\
\hline
Nonnegativity & $C(\hat{\varrho})\geq 0$, in general. \\[1ex]
Monotonicity & {\small $C$ does not increase under the action of incoherent operations.} \\[1ex]
Strong monotonicity & {\small $C$ does not increase on average under selective incoherent operations.} \\[1ex]
Convexity & {\small $C$ is a convex function of the state.}  \\[1ex]
Uniqueness  & {\small For any pure states $\vert\psi\rangle$, $C$ takes the form: \linebreak $C(\vert\psi\rangle\langle\psi\vert)=S(\Delta[\vert\psi\rangle\langle\psi\vert])$.} \\[1ex]
Additivity & {\small $C$ is additive under tensor products.}  \\[1ex]
\hline
\end{tabular}
\end{table}

Now, let us have a glimpse on the order of the magnitude of the decoherence time in the case of our problem. By using the thermal de Broglie wavelength $\lambda_{dB}=1/\sqrt{2mk_BT}$, we could define a corresponding decoherence time as
\begin{align}
\label{dtime}
\tau_D=\frac{\Delta X^2}{\gamma \lambda_{dB}^2},
\end{align}
where $\Delta X$ is the dispersion in position space and
\begin{align}
\label{gamma}
\gamma=\gamma_0\omega\bar{n}\frac{r^2}{1+r^2},
\end{align}
with $r=\Lambda/\omega$ and $\bar{n}=(e^{\omega/k_BT}-1)^{-1}$. $\Lambda$ represents the cutoff frequency and $\bar{n}$ denotes the mean population of the environment based on the Temperature ($T$). In the case of ours, we introduced the states consisting of two $K^{+}$ ions and two water molecules each interacting with channel wall, with its own dephasing rate $\gamma$. In this case, one can show that the system evolution can be written in a single Lindblad equation with the sum of all individual dephasing rates as its dephasing rate. The value of the decoherence time in decoherence formalism extremely depends on the value of this dephasing rate (equation \eqref{dtime}).

We examined the calculation of the decoherence time at the body temperature. We placed the values of mass, gamma, and dispersion in position space in Equations \eqref{dtime} and \eqref{gamma}. We used the reduced mass of potassium and water molecule. The value of the dispersion greatly depends on the frequency of the particles and varies between $1\times 10^{-12}$ to $1\times 10^{-10}m$. Changes in the value of the dephasing rate drastically change the value of the decoherence time. In this regard, we sketched the effect of the dephasing rate and the particles frequencies.

\begin{figure}
\centering
\includegraphics[scale=0.4]{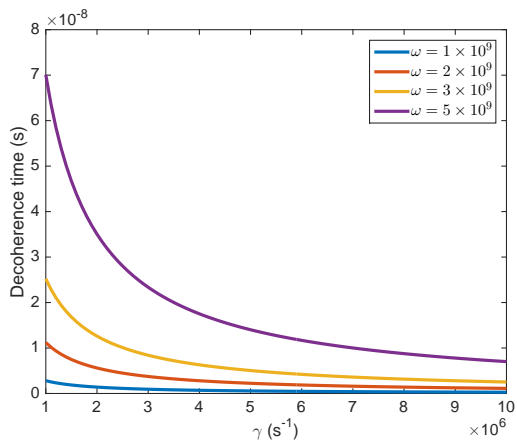}
\caption{Changes of the decoherence time as the dephasing rate increases, in a variety of system frequencies.} \label{dectime}
\end{figure}

As it is shown in Fig. \ref{dectime}, increasing in dephasing rate, decreases the decoherence time, on the other hand, an increase in system frequency also increases the decoherence time.  Accordingly, we considered the dephasing rate in the range of $\gamma=1\times10^6s^{-1}$ to $\gamma=1\times10^8s^{-1}$ , as reported in \cite{vaz}. In the mentioned range of the dephasing rate and with the system frequency in the range of $1\times10^9s^{-1}$ to $1\times10^{12}s^{-1}$, the decoherence time varies between $1\times10^{-10 }s$ to $1\times10^{-7} s$. If we consider other mechanisms in the absence of water molecules, our model can still predict the overall behavior of the system. Due to the change in frequency and reduced mass, the overall behavior of the system does not change, but the decoherence time coefficients and the slope of the graph may change. This is because we reduced the whole mechanism to a three-level process. Our model states are a three-step process. The decoherence time is not short for a particle to cross the channel, but definitely, it is not secure that the system remains coherent in the whole process. Therefore, we need to have an accurate study on the coherency of the system.

So far, many approaches have been proposed to quantify coherence. One of the most useful of them is the framework developed by Baumgratz \textit{et al.} \cite{lio,str}. Following this framework, a number of coherence quantifiers have been found. These include relative entropy of coherence, distillable coherence, the robustness of coherence, coherence concurrence, and coherence of formation \cite{bau,nap}. If the coherence quantifier satisfies all the postulates mentioned in TABLE \ref{table1}, it is called a coherence measure \cite{sol}. 
 
It has been proven that distillable coherence and relative entropy of coherence are equivalent and satisfy all the necessary conditions. Hence, they are considered as two basic coherence measures \cite{lio,hu}. Using these measures is mainly justified based on the physical intuition \cite{bau}. The distillable coherence is the optimal number of maximally coherent states which can be obtained from a state $\hat{\varrho}$ via incoherent operations \cite{str}. A simple expression for distillable coherence was given by Winter and Yang as follows \cite{yan}: 
\begin{equation}
\label{DistillableCoherence}
C_d(\hat{\varrho})=S(\Delta[\hat{\varrho}])-S(\hat{\varrho})
\end{equation}
\\where $\Delta[\hat{\varrho}]=\sum_{i=0}^{d-1}\vert i\rangle\langle i\vert\hat{\varrho}\vert i\rangle\langle i\vert$ is the dephasing operator and $S(\hat{\varrho})=-Tr[\varrho log_{2}\varrho] $ is the Von  Neumann entropy.

\begin{figure}
\centering
\includegraphics[scale=0.35]{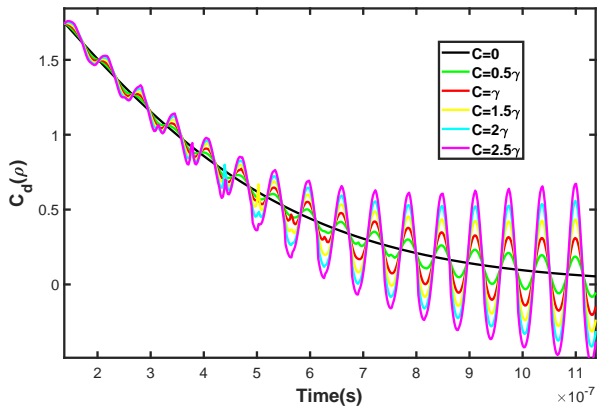}
\caption{The changes of the distillable coherence $C_d(\rho)$ versus time with $\omega=1\times10^8s^{-1}$, $\gamma=5\times10^6s^{-1}$ in a variety of hopping rates ($c$).} \label{dis}
\end{figure}

By studying the changes of distillable coherence, as we expect, the distillable coherence tends to zero as time passes by. In a hypothetical case with no hopping rate $c=0$, after the decoherence time the distillable coherence reaches zero $C_d(\rho)=0$  (see the black line in Fig. \ref{dis}). In the other hand, we sketched the distillable coherence against time in different hopping rates in Fig. \ref{dis}. The distillable coherence tends to zero yet, but interestingly, increasing the hopping rates causes the distillable coherence to oscillate. Therefore, the system remains coherent, even after the decoherence time. Moreover, a faster hopping rate results in increasing the amplitude of the oscillations and also more coherence.

\begin{figure}
\centering
\includegraphics[scale=0.45]{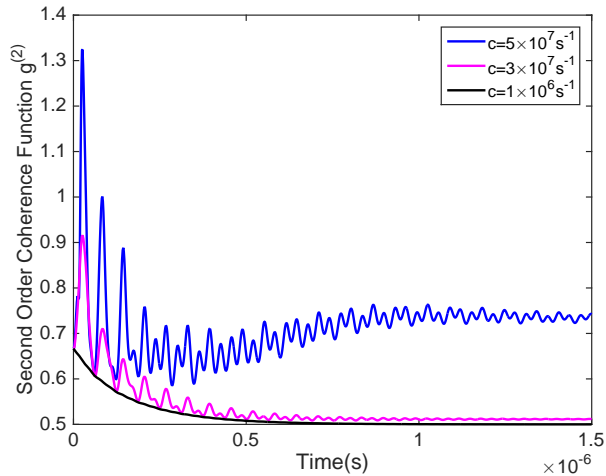}
\caption{The second order degree of coherence versus time in different hopping rates, with $\omega=1\times10^8s^{-1}$ and $\gamma=5\times10^6s^{-1}$.} \label{g2f}
\end{figure}

To support our findings, we calculate the second-order quantum coherence function  $g^{(2)}$(or the degree of coherence). At a fixed position $g^{(2)}$ depends only on the time difference $\tau$ and defines as
\begin{align}
\label{dc}
g^{(2)}(\tau)=\frac{\langle E^{(-)}(t)E^{(-)}(t+\tau)E^{(+)}(t+\tau)E^{(+)}(t)\rangle}{\langle E^{(-)}(t)E^{(+)}(t)\rangle\langle E^{(-)}(t+\tau)E^{(+)}(t+\tau)\rangle}
\end{align}
where,
\begin{align}
E^{(+)}=i(\frac{\hbar\omega_0}{2\epsilon_0V})^{1/2}\hat{S}_{-}e^{i(\bf{k.r}-\omega_0 t)}
\end{align}
is the positive frequency part of the field operator with wave vector ${\bf k}$, and $E^{(-)}$ is its conjugate. For a coherent state $\vert\alpha\rangle$ the degree of coherence is $g^{(2)}(\tau)=1$. Also, for a field in a single mode thermal-state it can be shown that $g^{(2)}(\tau)=2=1+\vert g^{(1)}(\tau)\vert^2$, which lies in the range of $1\leq g^{(2)}(\tau)\leq2$ and is just as in the classical case. On the other hand, in nonclassical systems such as numbers states the degree of coherence is in the range of $0<g^{(2)}(\tau)\leq1$. In the case of ours, for lower hopping rates $g^{(2)}(\tau)$ tends to $1/2$ after the decoherence time. By increasing the hopping rate ($c$), not only the oscillations appear, but also $g^{(2)}(\tau)$ increases in high hopping rates as we showed in Fig. \ref{g2f}. This observation confirms that a high throughput rate is a necessary requirement for an ion channel to stay coherent.

\section{Conclusion}

The objective of this paper is to give a clear answer to the paradox of the coexistence of a very high throughput rate of $K^+$ ions in KscA channels and their very high selectivity. In recent years, it has been suggested that quantum coherence is included in the process. However, the biological temperature and channels coupling with environment cause the system to lose coherence. 

In this regard, to investigate the problem, we studied a model that assume the system being in a superposition of three states: i) Potassium ions in the first and third sites and water molecules in the second and fourth sites. ii) Potassium ions in the second and fourth states with water molecules in the first and third sites. iii) A Potassium ion releases to the environment from the previous state. As the system interacts with the environment, it goes through a decoherence process. We studied the process using Lindblad master equation and solved the equations numerically.

The results shows that by measuring the system after the decoherence time, it almost always can be found in a state where release a Potassium ion to the environment. This observation is in accordance with the high throughput rate, but it does not necessitate quantum coherence in the system. Also, we discussed the decoherence time in the formalism of quantum decoherence in a variety of system frequencies and dephasing rates. Increasing the dephasing rate decreases the decoherence time, on the other hand, an increase in system frequency results a longer decoherence time. To study the system coherence directly, we used distillable coherence as a coherence measure and also calculated the second order coherence function. As we expect the distillable coherence tends to zero after the decoherence time. However, in high throughput rates, it oscillates from zero value and causes the system to remain coherent. Moreover, increasing the hopping rate increases the amplitude of the oscillations. The coherence function of the system tends to $1/2$ and by increasing the throughput rate, the value of the coherence function after the decoherence time increases but remains less that $1$. This observation clearly shows the point that the system is coherent in ion channels with high throughput rate as the systems that behaves classically have second order coherence functions greater than 1.

In this paper we successfully show that not only the high throughput rate is not in contrast with high selectivity, but also it is necessary for the system to remain coherent and act in quantum manner. In future works we concentrate on the topic that how the quantum coherence can act in the selectivity process.
\section*{Additional information}
Correspondence should be addressed to Afshin Shafiee [email: shafiee@sharif.edu].


\end{document}